\documentclass[twocolumn, times, tighten,twocolappendix]{aastex631}
\usepackage{graphicx,multirow,hyperref,url,color,xspace}
\usepackage{amsmath,amssymb}

\newcommand{\msun}{{\rm M}_{\sun}}

\newcommand{\gro}{{\textit{Compton Gamma Ray Observatory}}\xspace}
\newcommand{\swift}{{\textit{Neil Gehrels Swift}}\xspace}

\newcommand{\sax}{{\textit{Beppo\-SAX}}\xspace}

\newcommand{\g}{$\gamma$}

\newbox\grsign \setbox\grsign=\hbox{$>$} \newdimen\grdimen \grdimen=\ht\grsign
\newbox\simpropbox
\setbox\simpropbox=\hbox{\raise.5ex\hbox{$\propto$}\llap
 {\lower.5ex\hbox{$\sim$}}}\ht2=\grdimen\dp2=0pt

\defcitealias{Zdziarski22}{Z22}
\newcommand{\zdz}{{\citetalias{Zdziarski22}}\xspace}

\begin{document}

\title{The Composition and Power of the Jet of the Broad-Line Radio Galaxy 3C 120}
\shorttitle{The jet of 3C\,120}
\author[0000-0002-0333-2452]{Andrzej A. Zdziarski}
\affiliation{Nicolaus Copernicus Astronomical Center, Polish Academy of Sciences, Bartycka 18, PL-00-716 Warszawa, Poland; \href{mailto:aaz@camk.edu.pl}{aaz@camk.edu.pl}}
\author[0000-0002-0870-4569]{Dakalo G. Phuravhathu}
\affiliation{Centre for Space Research, North-West University, Potchefstroom 2520, South Africa}
\author[0000-0003-1667-7334]{Marek Sikora}
\affiliation{Nicolaus Copernicus Astronomical Center, Polish Academy of Sciences, Bartycka 18, PL-00-716 Warszawa, Poland; \href{mailto:aaz@camk.edu.pl}{aaz@camk.edu.pl}}
\author[0000-0002-8434-5692]{Markus B{\"o}ttcher}
\affiliation{Centre for Space Research, North-West University, Potchefstroom 2520, South Africa}
\author[0000-0002-9875-7436]{James O. Chibueze}
\affiliation{Centre for Space Research, North-West University, Potchefstroom 2520, South Africa}
\affiliation{Department of Physics and Astronomy, Faculty of Physical Sciences,  University of Nigeria, Carver Building, 1 University Road,
Nsukka 410001, Nigeria}

\shortauthors{Zdziarski et al.}

\begin{abstract}
We calculated the electron-positron pair production rate at the base of the jet of 3C 120 by collisions of photons from the hot accretion flow using the measurement of its average soft gamma-ray spectrum by Compton Gamma Ray Observatory. We found that this rate approximately equals the flow rate of the leptons emitting the observed synchrotron radio-to-IR spectrum of the jet core, calculated using the extended jet model following Blandford \& K{\"o}nigl. This coincidence shows the jet composition is likely to be pair-dominated. We then calculated the jet power in the bulk motion of ions, and found it greatly exceeds that achievable by the magnetically arrested disk scenario for the maximum black hole spin unless the jet contains mostly pairs. Next, we found that the magnetic flux through the synchrotron-emitting jet equals the maximum poloidal flux that can thread the black hole. Finally, we compared two estimates of the magnetization parameter at the onset of the synchrotron emission and found they are in agreement only if pairs dominate the jet content.
\end{abstract}

\section{Introduction}
\label{intro}

We consider possible mechanisms to produce electron-positron (e$^\pm$) pairs at the base of relativistic jets. There are a lot of indications that extragalactic jets contain e$^\pm$ pairs dominating their content by number (e.g., \citealt{Ghisellini12, Pjanka17, Snios18, Sikora20, Liodakis22}). However, it is not clear what the mechanism producing those pairs is. 

One possibility is pair production in the ergosphere of a rotating black hole (BH; \citealt{BZ77}). If the charge density there is low enough, a strong electric field forms a `spark gap', where electrons are accelerated to relativistic energies. These emit energetic \g-rays by Compton-scattering ambient soft photons. The \g-rays in turn produce e$^\pm$ pairs in collisions with ambient photons. However, the electric field will be screened when the charge density in the ergosphere becomes higher than the Goldreich-Julian \citep{GJ69} density, $n_{\rm GJ}=\Omega B/2\pi e c$, where $\Omega$ is the BH angular velocity, $B$ is the strength of the magnetic field of the magnetosphere and $e$ is the electron charge. The pairs are produced above the gap (see fig.\ 1 in \citealt{Levinson11}) and thus the density there can exceed $n_{\rm GJ}$ by some factor, which \citet{Levinson11} estimate to be $\lesssim\! 10^3$. As found to follow from that estimate, this mechanism can be be efficient in low-luminosity sources \citep{Levinson11, Moscibrodzka11}. On the other hand, the electron density close to the BH in blazars has been estimated to be much larger, $\sim\! (10^{10}$--$10^{13})n_{\rm GJ}$ \citep{Nokhrina15}. Here we accounted for the synchrotron-emitting electrons only, without counting any background electrons, which \citet{Nokhrina15} also included. Even without those electrons, it appears unlikely that such high electron densities can be obtained by this process.  

\begin{figure}
\centerline{
\includegraphics[width=7.cm]{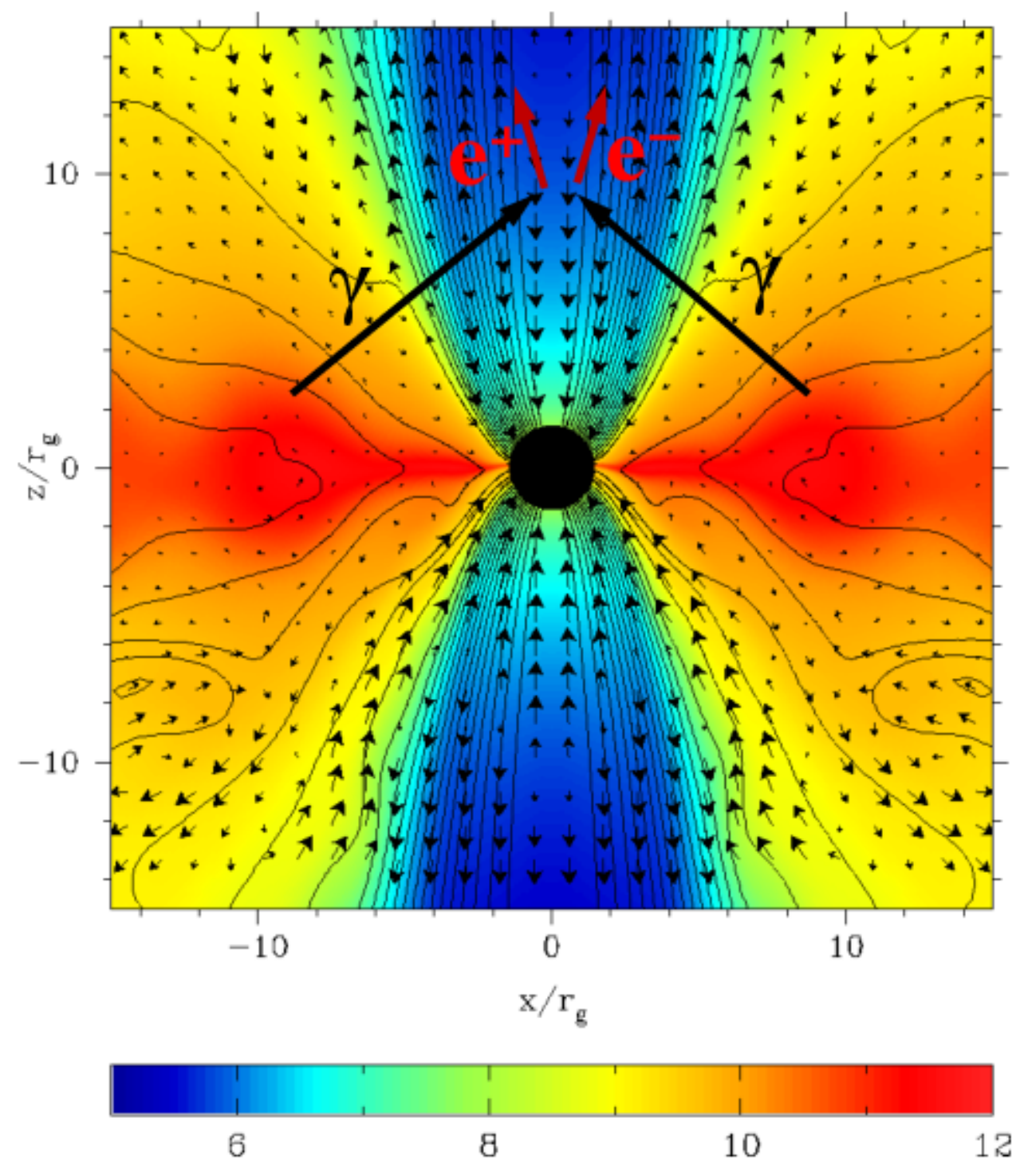}}
  \caption{A sketch of of the pair-producing geometry based on fig.\ 2a of \citet{Barkov08}, which shows the result of their GRMHD simulation for strongly magnetized accretion on a BH with the mass of $3\msun$ and the spin parameter of $a_*=0.9$. The contours show the magnetic field lines, the arrows show the velocity field, and the color contours show the baryonic rest mass density, $\rho$, in a logarithmic scale in $\log_{10}\rho$ in units of g\,cm$^{-3}$. The jet base, shown in blue, has a very low matter density. We show e$^\pm$ pair production within the jet base with the arrows representing pair-producing photons and the pairs. 
}\label{pairs}
\end{figure}

An alternative is pair production within the jet base via collisions of energetic photons originating in the accretion flow \citep{Henri91, B99_pairs, Levinson11, Aharonian17, Sikora20}. Figure \ref{pairs} presents an illustration of the assumed geometry (see also fig.\ 3 in \citealt{Zdziarski22}, hereafter \zdz). The rate of the pairs produced at the base can be compared with the flow rate of non-thermal relativistic electrons emitting synchrotron emission far downstream in the jet. Such a comparison was done for the microquasar MAXI J1820+070, where these two rates were found to be of the same order (\zdz). 

The ability of loading jets by pairs at their bases was also studied for AGNs using blazar and/or core-shift models combined with data on their radio lobes \citep{Kang14,Sikora16, Pjanka17}. However, this can be tested in some AGNs more directly,  in a similar way as in the case of the microquasar MAXI J1820+070, i.e., by using  radio-IR spectra of compact jets and hard X-ray/soft \g-ray spectra of their accretion disk coronae. This is usually not possible in blazars, where the beamed jet emission dominates over the X-rays from accretion. A suitable class of objects is broad-line radio galaxies, where the jet is viewed from a side. We have found the broad-line radio galaxy 3C\,120 to be well suited for such calculations.

This is a well-studied, strong and prominent jet source \citep{Marscher02_3C120}, but at the same time its X-ray emission is dominated by its accretion flow, as evidenced by the detection of strong reflection features \citep{ZG01,Lohfink13}. The model of the broad-band jet spectrum from 3C\,120 by \citet{Janiak16} also implies a weak jet contribution to the X-rays. The source was observed about 10 times by the Oriented Scintillation Spectroscopy Experiment (OSSE) detector \citep{Johnson93} onboard of {\it Compton Gamma Ray Observatory}, whose average spectrum was published by \citet{Wozniak98}, and compared to a number of X-ray observations at lower energies in \citet{ZG01}. Apart from a moderate difference in the normalization, the OSSE spectrum (starting at 50\,keV) was in a good agreement with that from \sax (extending up to $\approx$100\,keV), but showed a power-law extension at high energies, with the detection up to $\approx$400\,keV. For the synchrotron spectrum, we use the spectral measurements from radio to IR listed in \citet{Giommi12}.

The redshift to 3C\,120 is $z_{\rm r}=0.033$, which implies the luminosity distance of $D\approx 141$\,Mpc (for $H_0=72$\,km\,s$^{-1}$\,Mpc$^{-1}$, $\Omega_\Lambda=0.7$, $\Omega_{\rm M}=0.3$). The latest determination of its BH mass is $M\approx 6.92^{+2.63}_{-2.45}\times 10^7\msun$ \citep{Grier17}. It corresponds to the Eddington luminosity of $L_{\rm E}\approx 1.0\times 10^{46}$\,erg\,s$^{-1}$ assuming the H mass fraction of $X=0.7$, and to the gravitational radius of $R_{\rm g}\equiv GM/c^2\approx 1.0 \times 10^{13}$\,cm. Its accretion luminosity was estimated by \citet{Ogle05}, \citet{Kataoka11}, \citet{ZSPT15} and \citet{Janiak16} as $L_{\rm accr}\approx 1.5$, 0.8, 0.8 and $1.7\times 10^{45}$\,erg\,s$^{-1}$, respectively, which correspond to the Eddington ratio of 0.08--0.17. Given this relatively large ratio, the accretion efficiency, $\epsilon$, is implied to be large, likely in the 0.1--0.3 range. The jet inclination was measured as $i\approx 20.5\pm 1.8\degr$ \citep{Jorstad05}, $i\approx 16\degr$ \citep{Agudo12}, $i\approx 10.4\pm 2.3\degr$ \citep{Jorstad17} and $i\approx 18.7\pm 0.5\degr$ \citep{Pushkarev17}. The average value of the jet Lorentz factor was measured as $\Gamma\approx 5.3\pm 1.2$ \citep{Jorstad05}, $\Gamma\approx 10.7\pm 2.4$ \citep{Jorstad17} and $\Gamma\approx 7.9\pm 0.6$ \citep{Pushkarev17}. The jet half-opening angle was estimated as $\Theta\approx 3.8\pm 1.3\degr$ \citep{Jorstad05}, $\Theta\approx 1.2\pm 0.5\degr$ \citep{Jorstad17} and $\Theta\approx 3.1\pm 0.1\degr$ \citep{Pushkarev17}. Based on the above estimates, we will use in this work the following values,
\begin{align}
&i\approx 15\pm 5\degr,\quad \Gamma\approx 7.5\pm 2.5, \quad \Theta\approx 2\pm 1\degr,\label{pars}\\
&L_{\rm accr}\approx 1.0\pm 0.2 \times 10^{45}{\rm erg/s},\quad \epsilon\approx 0.2\pm 0.1.
\label{pars2}
\end{align}

\section{The spectra and model}
\label{model}

\begin{figure}
\centerline{\includegraphics[width=\columnwidth]{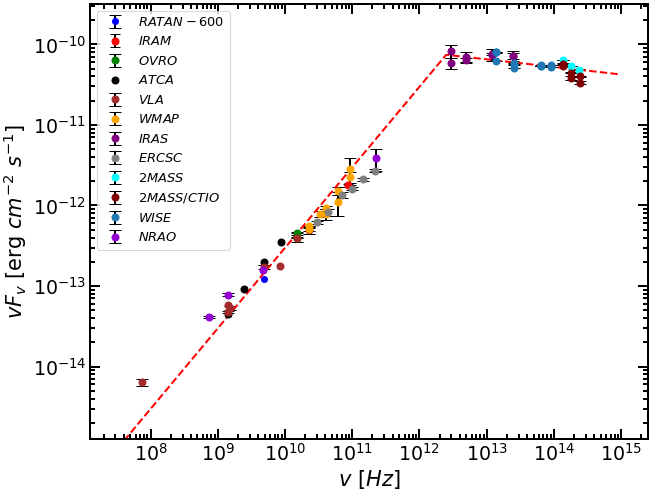}}
\caption{The radio to IR spectrum of the core of 3C\,120 from various measurements compiled in \citet{Giommi12}. The red dashed line gives a fit of the broken power law approximation to the partially synchrotron self-absorbed jet spectrum, Equation (\ref{solution}).
}
 \label{spectrum}
 \end{figure}

Figure \ref{spectrum} shows the radio to IR spectrum of the core of 3C\,120, representing the inner and steady part of the jet. While the shown measurements were not simultaneous, we see relatively little scatter among them, indicating that the core compact jet is indeed stable. Its radio to IR spectrum has a power-law part approximately flat in $F_\nu$ (below $\sim\! 10^{12}$\,Hz), followed by a significantly steeper power law at higher frequencies. This implies the presence of a break frequency in-between. The low-$\nu$ part is usually interpreted as coming from jet synchrotron emission being partially self-absorbed \citep{BK79, Konigl81}, while the part above the break frequency comes from synchrotron emission being optically thin in the entire emitting part of the jet. There is a possible contribution in the IR from the molecular torus. However, it appears to be negligibly small in 3C\,120, which we have inferred based on the quasar template in \citet{Richards06}.

The radio-to-submm part of the spectrum forms a hard power law, which we fitted\footnote{Some of the fluxes we compiled from the literature had no measurement errors, and for the $\chi^2$ fit (weighted by inverse squares of the errors) we assumed them to equal 5\% of the flux.} as $F_\nu\propto \nu^\alpha$ with $\alpha\approx -0.20\pm 0.04$ (where the uncertainty is $1\sigma$). This fits well its entire range of 74\,MHz--226\,GHz. However, an extrapolation of this power law to the IR band intersects the softer power law at a frequency a factor of a few above its minimum measured energy. This shows the overall spectrum is more complex than that implied by the model of \citet{BK79}. Namely, the $\alpha\approx -0.20$ radio-to-submm power law has to change to a harder one in order to connect to the lowest IR point. Such changes are related to the rates of the changes of the density of the relativistic electrons and of the magnetic field strength, see, e.g., \citet{Konigl81}, \citet{Zdziarski19b}. However, our goal here is to estimate the rate of the electron flow at the onset of the emitting region (with $z\propto \nu^{-1}$) and we do not need to consider this complexity. Thus, we approximate the overall spectrum as a once-broken power law, with the two parts intersecting around the break frequency. The hard power law has $\alpha_{\rm thick}=0$, and the power law fitted to the IR part of the spectrum has $\alpha_{\rm thin}\approx -1.09\pm 0.02$. This is shown by the dashed line in Figure \ref{spectrum}, where we see it still provides a fair approximation to the entire spectrum. 

In our modelling, we follow the formulation of the partially synchrotron self-absorbed conical jet model given in \zdz, which we summarize in Appendix \ref{app} below. The break frequency, $\nu_0$, is defined as corresponding to a unit optical depth at the base of the emitting part, $z_0$. The broken-power law approximation is given by Equation (\ref{solution}). For 3C\,120, we find $\nu_0$ and $F_\nu$ at $\nu_0$, denoted hereafter as $F_{\nu_0}$, as
\begin{equation}
\nu_0\approx 3.15\times 10^{3}\,{\rm GHz},\quad F_{\nu_0}\approx 3.0\,{\rm Jy}.
\label{nuF}
\end{equation}
The hard index of $\alpha_{\rm thick}=0$ corresponds to the canonical \citep{BK79} power law dependencies of the electron number density and the magnetic field strength, $n(z,\gamma)= n_0(z/z_0)^{-a}\gamma^{-p}$ (where $\gamma$ is the electron Lorentz factor), $B(z)=B_0 (z/z_0)^{-b}$, respectively, with $b=a/2=1$. The soft index of $\alpha_{\rm thin}=-1.09$ corresponds to the power law index of the synchrotron-emitting electrons of $p=1-2\alpha_{\rm thin}\approx 3.18$. We also need to set the minimum and maximum values of $\gamma$, which we assume to be $\gamma_{\rm min}=2$ (approximately the minimum for which the relativistic formulae are valid) and $\gamma_{\rm max}=10^4$. These values give a low average electron kinetic energy of $\approx\!\! 2.7 m_{\rm e}c^2$, where $m_{\rm e}$ is the electron mass. Given the relative steepness of the electron distribution, all derived quantities weakly depend on $\gamma_{\rm max}$. Furthermore, we assume the ratio between the energy densities of the matter (kinetic only) and magnetic field of $\beta_{\rm eq}=1$, where we further assume the only contribution to the matter kinetic energy density is from the relativistic electrons, see Equation (\ref{betaeq}). We stress, however, that a contribution from ions is possible.

\begin{figure}
\centerline{\includegraphics[width=7cm]{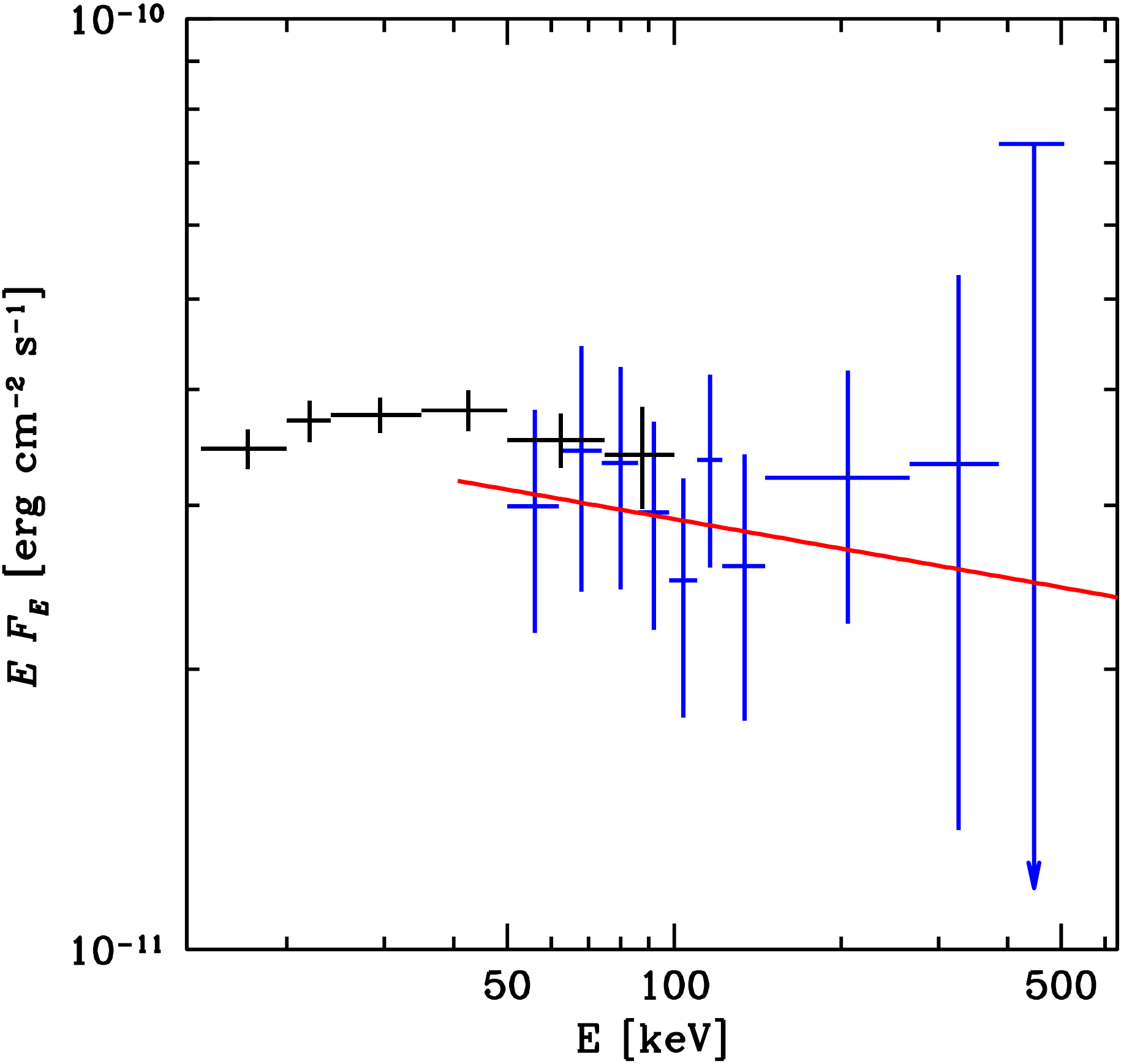}}
\caption{The blue symbols show the average 1992--97 \gro OSSE spectrum of 3C\,120 from \citet{Wozniak98}. The red line gives the best power-law fit in the 50--500\,keV range. For comparison, we show the 2004--2013 average spectrum \citep{Oh18} from the \swift BAT (fitted separately by a power law, not shown).
}
 \label{OSSE}
 \end{figure}

In order to estimate the uncertainty ranges of our derived quantities, we find their extreme values by varying the values of $i$, $\Gamma$, $\Theta$ and $L_{\rm accr}$ in the ranges given in Equations (\ref{pars}--\ref{pars2}). Some of those ranges are large, violating the assumptions behind the standard propagation of errors, and thus the above procedure gives more realistic, as well as conservative, error ranges. 

Knowing $\nu_0$, $F_{\nu_0}$, $p$ (from the observed spectrum), and $D$, $\Gamma$, $i$, $\Theta$ (from other observations, Section \ref{intro}), as well as assuming $\gamma_{\rm min}$, $\gamma_{\rm max}$, $\beta_{\rm eq}$, we can use standard formulae for the synchrotron emission and absorption and the relativistic transformations and determine the distance of the onset of the emission, $z_0$, and the magnetic field strength there, $B_0$, see Equations (\ref{z_0}) and (\ref{B_0}), respectively. For our estimated values, we find
\begin{align}
&z_0\approx 2.0_{-0.5}^{+1.3}\!\times\! 10^{16}\,{\rm cm} \approx 1.9_{-0.4}^{+1.3}\!\times\! 10^{3}R_{\rm g},\label{z0}\\
&B_0 \approx 47_{-17}^{+31}\,{\rm G}.
\label{B0}
\end{align}
The dependence of $z_0$ on $\beta_{\rm eq}$ is very weak, as $\propto \beta_{\rm eq}^{-0.05}$, and also the dependence on $\gamma_{\rm min}$ is weak. Our estimate of $z_0$ is consistent with the existing estimates of the location of blazar zones (where the bulk of the high-energy emission originates), which are found at distances between those of the broad-line regions and the molecular torii, $\sim\!\! 10^3$--$10^5 R_{\rm g}$ (e.g., \citealt{MS16}). 

We can compare the above estimate of $B_0(z_0)$ with that using measurements of the core shift,  which is an angular displacement of the position of the radio core between two frequencies \citep{Lobanov98}. It appears that the only such measurements for 3C\,120 are those of the angular shift between 2.3 and 8.6\,GHz of $1.196\pm 0.018$ and $0.956\pm 0.016$\,mas by \citet{Kovalev08}. To estimate the magnetic field from them, we use eq.\ (7) of \citet{ZSPT15}. Using the range of the measured core shift and the assumed $\Gamma$, $i$ and $\Theta$, we find $B_0(z_0)\approx 27$ and 23\,G, respectively, with the total uncertainty range of 18--34\,G, i.e., about 50\% of the values estimated from the break frequency and the flux. Since the two sets of values are estimated based on frequencies separated by a factor of $\sim\! 10^2$, we consider them to be in a relatively good agreement.

Given the values of $B_0$ and $\nu_0$, we can estimate the electron Lorentz factor corresponding to the bulk of the emission in the partially self-absorbed part, see eq.\ (16) of \zdz, 
\begin{equation}
\gamma_{\rm bulk}\approx\left(\frac{B_{\rm cr}}{B_0}\frac{h\nu_0}{\delta m_{\rm e}c^2}\right)^{1/2}\approx 90^{+10}_{-10},
\label{gamma_bulk}
\end{equation}
where $\delta$ is the Doppler factor, $B_{\rm cr}$ is the critical magnetic field strength, and the numerical value is for our determined parameters. For the validity of the assumed model, $\gamma_{\rm min}$ needs to be a factor of at least a few lower than $\gamma_{\rm bulk}$. This shows $\gamma_{\rm min}$ needs to be relatively low, but it could still be larger than our assumed $\gamma_{\rm min}=2$.

Next, we can calculate the rate of the flow of electrons through the jet, $\dot N_{\rm e}$, which is given by the relativistic number flux, see Equation (\ref{Ndot_e}). Substituting our assumed values and obtained solutions, we find
\begin{equation}
\dot N_{\rm e}\approx 2.7_{-2.3}^{+12}\times 10^{49}\,{\rm s}^{-1}\propto
\Theta^{\frac{8+2 p}{13+2 p}}\gamma_{\rm min}^{\frac{7-12 p}{13+2 p}},
\label{Ndot}
\end{equation}
which is an increasing function of $\Gamma$, $\Theta$ and $i$. The dependence on $\gamma_{\rm min}$ is approximate, valid for $\gamma_{\rm min}\gg 1$. 

We then compare the rate of Equation (\ref{Ndot}) to an estimate of the pair production rate within the jet base by collisions of photons from the accretion flow. We use the average OSSE spectrum obtained by \citet{Wozniak98}, shown on Figure \ref{OSSE}. For comparison, we also show the average spectrum\footnote{\url{https://swift.gsfc.nasa.gov/results/bs105mon/226}} from \swift Burst Alert Telescope (BAT; \citealt{Oh18}). Apart from a slightly higher normalization, its part at $\gtrsim$50\,keV is compatible with the average from the OSSE. We have fitted the OSSE spectrum by a power law, shown by the red line. We find $\alpha_{\rm X}\approx -1.11^{+0.27}_{-0.26}$ and the normalization of $EF_E(511\,{\rm keV})\approx 0.015^{+0.07}_{-0.05}$\,keV/(cm$^2$\,s), where the uncertainties are $1\sigma$ (but $\alpha_{\rm X}$ and $EF_E$ are strongly correlated). This gives us the density of photons within the jet base, Equation (\ref{n1}). We then calculate the rate of pair production by photons with such a power law, which also depends on the assumed characteristic sizes of the hot plasma, $R_{\rm hot}$, and the jet base, $R_{\rm jet}$, see Equation (\ref{pprate}). From it, we obtain, 
\begin{equation}
2\dot N_+\approx 3.9^{+6.3}_{-2.8}\times 10^{49}\left(\frac{R_{\rm hot}}{10 R_{\rm g}}\right)^{-1}\left(\frac{R_{\rm jet}}{R_{\rm hot}}\right)^2 {\rm s}^{-1}.
\label{Nplus}
\end{equation}
(The given uncertainty is for the $1\sigma$ joint $\alpha_{\rm X}$--$EF_E(511\,{\rm keV})$ error contour.) This pair production rate is then balanced by the sum of the rates of pair annihilation and pair advection. The equilibrium Thomson optical depth of the pairs is of the order of unity. Still, the advection downstream along the jet is likely to dominate since the pairs are produced with a substantial net momentum along the jet axis \citep{B99_pairs}. 

We see that the two rates are similar. This is a remarkable coincidence since the two numbers have been derived with very different physics. This shows that pair production by accretion photons is fully capable to provide most of the leptons responsible for the jet synchrotron emission. The same conclusion was reached for the hard spectral state of the accreting X-ray binary MAXI J1820+070. 

Another argument for the presence of e$^\pm$ pairs in the source is given by a calculation of the jet power, $P_{\rm j}$. The total jet power in the framework of the theoretically predicted \citep{BK74,BK76,Narayan03} and numerically confirmed \citep{Punsly09,Tchekhovskoy11,McKinney12} magnetically arrested disk (MAD) scenario is
\begin{equation}
P_{\rm j}\approx 1.3\left(\phi_{\rm BH}/50\right)^2 a_*^2\dot M c^2,
\label{P_tot}
\end{equation}
where $\dot M$ is the accretion rate, $a_*$ is the BH spin parameter, and $\phi_{\rm BH}$ is a dimensionless magnetic flux threading the BH on one hemisphere related to the dimensional flux by $\Phi_{\rm BH}=\phi_{\rm BH}(\dot M c)^{1/2} R_{\rm g}$. GRMHD simulations have shown that $\phi_{\rm BH}\lesssim 50 h_{\rm 0.3}^{1/2}$, where $h_{0.3}$ is a dimensionless disk half-thickness defined by $H_{\rm disk}=R_{\rm disk} 0.3 h_{\rm 0.3}$ \citep{Davis20}. Noting that $a_*<1$ and $h_{\rm 0.3}\lesssim 1$, $P_{\rm j}\lesssim \dot M c^2\equiv L_{\rm accr}/\epsilon$, which is also a general, model-independent, constraint. In 3C\,120, we find 
\begin{equation}
P_{\rm j}\lesssim 5\pm 1\times 10^{45}\left(\epsilon/0.2\right)^{-1}{\rm erg\,s}^{-1}.
\label{Pj}
\end{equation}

In our model, we can then calculate the jet power in the magnetic field and electrons, $P_{B{\rm e}}$, using Equation (\ref{PBe}),
\begin{equation}
P_{B{\rm e}}\approx 1.5_{-1.3}^{+10.2}(1/2+\beta_{\rm eq}/3)\times 10^{45} {\rm erg\,s}^{-1}.
\label{Peb}
\end{equation}
The power in cold ions associated with the electrons not in e$^\pm$ pairs is given by Equation (\ref{P_ion}). We obtain
\begin{equation}
P_{\rm i}\approx 3.1^{+19.6}_{-2.8}\times 10^{47}\left(1-2 n_+/n_{\rm e}\right) {\rm erg\,s}^{-1},
\label{Pi}
\end{equation}
where $n_+$ and $n_{\rm e}$ are the densities of the positrons and of all leptons, respectively, and in steady state $n_+/n_{\rm e}\approx \dot N_+/\dot N_{\rm e}$. Both $P_{B{\rm e}}$ and $P_{\rm i}$ increase with $\Gamma$, $i$ and $\Theta$, while $P_{B{\rm e}}/P_{\rm i}\approx 0.005/(1-2 n_+/n_{\rm e})$ (for $\beta_{\rm eq}\approx 1$) remains approximately constant. Thus, the power in the electrons and $B$ is negligible compared to that in the ions unless the jet consists of almost pure pair plasma. Then, $P_{\rm i}/P_{\rm j}\gtrsim 62^{+860}_{-60}(1-2 n_+/n_{\rm e})$. Thus, the presence of pairs is required by the physical requirement of $P_{\rm i}\leq P_{\rm j}$, implying $1-2 n_+/n_{\rm e}\ll 1$ (i.e., a pair-dominated jet) in most of the parameter space. On the other hand, $P_{\rm i}$ is approximately $\propto\gamma_{\rm min}^{(7-12 p)/(13+2 p)}$, and thus the pair abundance, $2 n_+/n_{\rm e}$, can be close to null if $\gamma_{\rm min}\gg 2$.  

The above comparison also implies the jet power of 3C\,120 can be near the maximal possible power. In the case of the jet extracting the BH rotational power, we can test this conjecture by comparing the magnetic flux threading the BH with that estimated from the observed synchrotron emission downstream in the jet at a $z\geq z_0$. The latter was determined for a sample of radio loud AGNs in \citet{Zamaninasab14} and \citet{ZSPT15}. We use eq.\ (21) in \citet{ZSPT15} for the quantities at $z_0$,
\begin{equation}
\Phi_{\rm j}=2^{3/2}\pi R_{\rm H}s z_0 B_0(1+\sigma_0)^{1/2}/(\ell a_*),
\label{phi_jet}
\end{equation}
where $R_{\rm H}=[1+(1-a_*^2)^{1/2}] R_{\rm g}$ is the BH horizon radius, $\ell\la 0.5$ is the ratio of the  angular frequencies of the magnetic field lines dragged around the BH and the BH, $s\lesssim 1$ is defined by 
\begin{equation}
\sigma_0\equiv (\Theta\Gamma/s)^{2}
\label{s}
\end{equation}
\citep{Komissarov09,Tchekhovskoy09}, and $\sigma_0$ is the magnetization parameter at $z_0$. In our estimates below, we adopt $a_*=1$ and $\ell=0.5$. We find
\begin{align}
&\Phi_{\rm BH}\approx 2.1\pm 0.2(\phi_{\rm BH}/50)\left(\epsilon/0.2\right)^{-1/2} 10^{32} {\rm G\,cm}^2,
\label{Phi_BH}\\
&\Phi_{\rm j}\approx 1.7_{-0.9}^{+3.0}s(1+\sigma_0)^{1/2} 10^{32}\,{\rm G\,cm}^{2}.
\label{Phi_j}
\end{align}
We thus find here another remarkable coincidence (in agreement with that of \citealt{Zamaninasab14} and \citealt{ZSPT15}), of the two values of the magnetic flux fully compatible with being identical for $\phi_{\rm BH}\sim 50$ and $\sigma_0\lesssim 1$, expected at $z\gtrsim z_0$. Thus, the jet power can be maximal, corresponding to a jet ejected from a system with a MAD accretion flow and a maximally rotating BH.

Now we consider the value of the magnetization parameter, $\sigma_0$. By definition,
\begin{equation}
\sigma_0 \equiv \frac{B_0^2/4\pi}{\eta u_{\rm k,0}+\rho_0 c^2}\approx \frac{B_0^2/4\pi}{n_0 f_N \mu_{\rm e}m_{\rm p}c^2 (1-2 n_+/n_{\rm e})},
\label{sigma}
\end{equation}
where $\eta$ is the particle adiabatic index, $u_{\rm k,0}$ is the kinetic energy density of particles at $z_0$, $\rho_0 c^2$ is the rest-energy density at $z_0$, $f_N$ is given by Equation (\ref{fe_fn}), $\mu_{\rm e}=2/(1+X)$ is the mean electron molecular weight, and $m_{\rm p}$ is the proton mass. In our case, $u_{\rm k,0}\ll \rho_0 c^2$, e.g., 1/600 of the latter for our default parameters and in the absence of pairs, which in turn leads to the approximate equality on the right-hand side. For our assumed parameters, we find using Equations (\ref{sigma}) and (\ref{s}),
\begin{align}
\sigma_0 \approx \frac{2.5\times 10^{-3}}{\beta_{\rm eq}(1-2 n_+/n_{\rm e})},
\label{sigma1}\\
\sigma_0\approx 6.9^{+20.5}_{-6.1}s^{-2}\times 10^{-2},\label{sigma2}
\end{align}
respectively, where the former is independent of $\Gamma$, $i$ and $\Theta$. Setting the two values of $\sigma_0$ equal implies $(1-2 n_+/n_{\rm e})\beta_{\rm eq}\ll s^2$, i.e., a pair dominated plasma for most of the parameter space. For our default values, $2 n_+/n_{\rm e}\approx 0.96$, i.e., there are about 25 times more leptons in pairs than in the ionization electrons. We also note that our estimated values of $\sigma_0$ are $\ll$1.

\section{Summary and Discussion}
\label{summary}

We have found that the composition of the jet in 3C\,120 is most likely dominated by e$^\pm$ pairs. First, a sufficient amount of pairs can be produced by collisions of hard X-rays/soft \g-rays emitted by the accretion flow, as indicated by the estimated rate of pair production being approximately equal to the rate of the flow of synchrotron-emitting electrons downstream in the jet. Second, the high pair content is indicated by the calculation of the jet power, which is dominated by that in the bulk motion of the ions. If the ion density directly corresponds to that of the synchrotron-emitting electrons, the jet power would greatly exceed its maximum possible value, $\dot M c^2$, but it can be $\lesssim \dot M c^2$ if most of the emitting leptons are those in e$^\pm$ pairs. Third, the estimates of the magnetization parameter from the flow parameters and from the product of $\Theta$ and $\Gamma$ agree only at a high pair content. However, apart from the implied result that the pair content, $2 n_+/n_{\rm e}$ is $\approx$1, we cannot determine precisely its value because the jet parameters for this source are relatively loosely constrained.

In our calculations, we have assumed the jet is conical in the dissipation zone, at $z\geq z_0$, and we have estimated $z_0$ to equal a few times $10^{16}$\,cm. On the other hand, \citet{Kovalev20} found that the jet in 3C\,120 becomes conical only at distances about 100 times larger. A departure from the conical shape would, however, affect mostly the value of $z_0$. On the other hand, the parameters of the flow are determined by the jet radius, i.e., $\Theta z_0$, which is sensitive mostly to the local synchrotron emission and absorption, and are only weakly affected by the jet shape. Furthermore, we have found the radio-to-submm spectrum forms an almost pure power law at 74\,MHz--226\,GHz, which indicates that the shape of the jet in the emitting range, which starts relatively close to $z_0$ (since $z\propto \nu^{-1}$ and 226\,GHz is $\sim 0.1\nu_0$), is most likely conical and cannot have a substantial curvature. 

Another uncertainty in the assumptions regards the low-energy end of the electron distribution. On one hand, the value of $\gamma_{\rm min}$ is uncertain and can be higher than our assumed $\gamma_{\rm min}=2$. In this case, it will be even easier for the pair production at the jet base to produce enough leptons for the synchrotron emission, while it would reduce the jet power. On the other hand, we expect the presence of a quasi-thermal electron distribution below $\gamma_{\rm min}$ (regardless of its value), which would increase the rate of the electron flow, $\dot N_{\rm e}$. Depending on the relative fraction of the electrons in the quasi-thermal part, the value of $\gamma_{\rm min}$ and other jet parameters, this may result in the pair production rate being not sufficient to account for the full flow. Such a quasi-thermal distribution is expected in the presence of shock acceleration (e.g., \citealt{Sironi09, Sironi15b, Crumley19}). For acceleration via magnetic field reconnection, simulations performed for $\sigma\geq 1$ show the quasi-thermal part to be modest, with the lepton number still dominated by the power-law part (e.g., \citealt{Comisso18,Comisso19, Petropoulou19}). Furthermore, the presence of substantial excesses of cold electrons in jets of radio-loud AGNs (hence also in 3C\,120) is ruled out by the observed absence of soft X-ray excesses in spectra of blazars. Such an excess would result from Comptonization of external UV radiation by the cold electrons in a relativistic jet \citep{Sikora97}.

\section*{Acknowledgements}

We thank G. Henri, A. Levinson, P.-O.\ Petrucci and T. Savolainen for valuable discussions, the referee for a valuable suggestion, M. Barkov for the permission to use his figure in our illustration of the effect of pair production, and P. Lubi{\'n}ski and N. {\.Z}ywucka for help with software. We acknowledge support from the Polish National Science Center under the grants 2015/18/A/ST9/00746 and 2019/35/B/ST9/03944. The work of M.B. was supported in part by the National Research Foundation\footnote{Any opinion, finding and conclusion or recommendation expressed in this material is that of the authors, and the NRF does not accept any liability in this regard.} of South Africa (DSI/NRF SARChI Programme, Grant No.\ 64789). 

\appendix
\section{Formulae}
\label{app}

In order to make this work self-contained, we present the formulae from \zdz and \citet{Zdziarski21c} that are used here. In \zdz, they were given for arbitrary power-law dependencies of $n(z)$ and $B(z)$, while here we give them for the canonical case of $b=a/2=1$. 

We rewrite eqs.\ (12--13) of \zdz as
\begin{align}
F_\nu \simeq F_{\nu_0}
\min\left[ 1, \frac{5(\nu/\nu_0)^{\frac{1-p}{2}}}{(p-1)\Gamma\left(\frac{p-1}{p+4}\right)} \right],\label{solution}\\
\tau(\nu,z)=\left[(\nu/\nu_0)(z/z_0)\right]^{-(p+4)/2},\label{tau}
\end{align}
which give a broken power-law aproximation to the synchrotron spectrum and the self-absorption optical depth, respectively. Here $\nu_0$ is defined by $\tau(\nu_0,z_0)=1$, i.e., the jet has $\tau<1$ at $\nu>\nu_0$ in the entire synchrotron-emitting part, $z\geq z_0$, and the two power laws intersect at a $\nu<\nu_0$. Note that $F_{\nu_0}\equiv (2/5) \Gamma\left[(p-1)/(p+4)\right] F_0$, where $F_0$ is defined by eq.\ (7) of \zdz. We then use eqs.\ (A2) and (A4) of \zdz, which give the solutions for $z_0$ and $B_0$, respectively,
\begin{align}
&z_0=\frac{1}{\delta^{\frac{4+p}{13+2 p}} \nu_0} \left[\frac{c (1+k_{\rm i})(f_E-f_N)}{2\pi \beta_{\rm eq} }\right]^{\frac{1}{13+2 p}} \times \nonumber\\
&\left[\frac{C_2(p)}{\sin i}\right]^{\frac{5+p}{13+2 p}}\!\! 
 \left[\frac{5 F_{\nu_0} D^2}{m_{\rm e}C_1(p) \Gamma\left(\frac{p-1}{p+4}\right)}\right]^{\frac{6+p}{13+2 p}}\times \label{z_0}\\
&\left(\frac{3}{\pi\tan\Theta}\right)^{\frac{7+p}{13+2 p}} (1+z_{\rm r})^{-\frac{19+3p}{13+2 p}},\nonumber
\end{align}
\begin{align}
&B_0= \left[\frac{3 C_1(p) \Gamma\left(\frac{p-1}{p+4}\right)(1+k_{\rm i})^2 (f_E-f_N)^2\sin^3 i}{5 C_2(p)^3\beta_{\rm eq}^2 D^2 F_{\nu_0}\tan\Theta }\right]^{\frac{2}{13+2 p}} \nonumber\\
&\times \frac{\nu_0 \pi^{\frac{7+2 p}{13+2 p}}  2^{\frac{9+2 p}{13+2 p}} c^{\frac{17+2 p}{13+2 p}} m_{\rm e}^{\frac{15+2 p}{13+2 p}} (1+z_{\rm r})^{\frac{15+2 p}{13+2 p}}}{e \delta^{\frac{3+2 p}{13+2 p}}},
\label{B_0}
\end{align}
where $e$ is the electron charge, $C_{1,2}$ are constants defined in eqs.\ (8) and (9) of \zdz, respectively,
\begin{align}
&\beta_{\rm eq}={n_0 m_{\rm e} c^2 (1+k_{\rm i})(f_E- f_N)\over B_0^2/8\pi},\label{betaeq}\\
f_E\equiv& \begin{cases} {\gamma_{\rm max}^{2-p}-\gamma_{\rm min}^{2-p}\over 2-p}, &p\neq 2;\cr
\ln {\gamma_{\rm max}\over \gamma_{\rm min}},& p=2,\cr
\end{cases}\quad
f_N\equiv \frac{\gamma_{\rm min}^{1-p}-\gamma_{\rm max}^{1-p}}{p-1},
\label{fe_fn}
\end{align}
and $k_{\rm i}$ accounts for the kinetic energy density in particles other than the power-law electrons. 

The total lepton flow rate and the usable power in ions are
\begin{align}
\dot N_{\rm e} &\approx 2\pi n_0 f_N c\beta\Gamma (z0 \tan\Theta)^2,
\label{Ndot_e}\\
P_{\rm i}&=\mu_{\rm e} m_{\rm p}c^2 (\Gamma-1)\left(\dot N_{\rm e}- 2\dot N_+\right),\label{P_ion}
\end{align}
respectively, where $\beta$ is the dimensionless bulk velocity, and both the jet and counterjet are taken into account, see eqs.\ (25--26) of \zdz. The power in the relativistic electrons and magnetic fields is, see eq.\ (23) of \zdz,
\begin{equation}
P_{B{\rm e}}=\left(\frac{1}{2}+\frac{\beta_{\rm eq}}{3}\right)\! c\beta(B_0 z_0 \Gamma \tan\Theta)^2 .
\label{PBe}
\end{equation}

We then provide the formalism for calculating the pair production rate, based on \citet{Zdziarski21c}. We modify their eq.\ (1) to calculate the differential photon density at 511\,keV, $n_1$, {\it above\/} the hot disk,
\begin{equation}
n_1\approx F_E(511\,{\rm keV})\frac{4\pi D^2}{2\pi R_{\rm hot}^2 c},
\label{n1}
\end{equation}
where we divided the total rate of the photon emission by the source area (including both sides), and both $F_E$ and $E$ have the same energy unit (e.g., keV). We then use eq.\ (3) of \citet{Zdziarski21c}, which uses the rate derived by \citet{Svensson87}, and approximate the pair-producing volume as two cylinders with the height $R_{\rm hot}$ and the characteristic radius of the jet, $R_{\rm jet}$ ($\leq R_{\rm hot}$), i.e., $V=2\pi R_{\rm jet}^2 R_{\rm hot}$. We then get the total pair production rate as
\begin{equation}
\dot N_+\approx 2\pi R_{\rm jet}^2 R_{\rm hot} n_1^2 \sigma_{\rm T} c\frac{\ln(E_{\rm c}/511\,{\rm keV})} {(1-\alpha_{\rm X})^{5/3} (2-\alpha_{\rm X})},
\label{pprate}
\end{equation}
where $\sigma_{\rm T}$ is the Thomson cross section, and $E_{\rm c}$ is the upper cutoff of the photon power law, which we assume $\approx$2\,MeV.

\bibliography{allbib}{}

\begin{thebibliography}{}
\expandafter\ifx\csname natexlab\endcsname\relax\def\natexlab#1{#1}\fi
\providecommand{\url}[1]{\href{#1}{#1}}
\providecommand{\dodoi}[1]{doi:~\href{http://doi.org/#1}{\nolinkurl{#1}}}
\providecommand{\doeprint}[1]{\href{http://ascl.net/#1}{\nolinkurl{http://ascl.net/#1}}}
\providecommand{\doarXiv}[1]{\href{https://arxiv.org/abs/#1}{\nolinkurl{https://arxiv.org/abs/#1}}}

\bibitem[{{Agudo} {et~al.}(2012){Agudo}, {G{\'o}mez}, {Casadio}, {Cawthorne},
  \& {Roca-Sogorb}}]{Agudo12}
{Agudo}, I., {G{\'o}mez}, J.~L., {Casadio}, C., {Cawthorne}, T.~V., \&
  {Roca-Sogorb}, M. 2012, \apj, 752, 92, \dodoi{10.1088/0004-637X/752/2/92}

\bibitem[{{Aharonian} {et~al.}(2017){Aharonian}, {Barkov}, \&
  {Khangulyan}}]{Aharonian17}
{Aharonian}, F.~A., {Barkov}, M.~V., \& {Khangulyan}, D. 2017, \apj, 841, 61,
  \dodoi{10.3847/1538-4357/aa7049}

\bibitem[{{Barkov} \& {Komissarov}(2008)}]{Barkov08}
{Barkov}, M.~V., \& {Komissarov}, S.~S. 2008, \mnras, 385, L28,
  \dodoi{10.1111/j.1745-3933.2008.00427.x}

\bibitem[{{Beloborodov}(1999)}]{B99_pairs}
{Beloborodov}, A.~M. 1999, \mnras, 305, 181,
  \dodoi{10.1046/j.1365-8711.1999.02384.x}

\bibitem[{{Bisnovatyi-Kogan} \& {Ruzmaikin}(1974)}]{BK74}
{Bisnovatyi-Kogan}, G.~S., \& {Ruzmaikin}, A.~A. 1974, \apss, 28, 45,
  \dodoi{10.1007/BF00642237}

\bibitem[{{Bisnovatyi-Kogan} \& {Ruzmaikin}(1976)}]{BK76}
---. 1976, \apss, 42, 401, \dodoi{10.1007/BF01225967}

\bibitem[{{Blandford} \& {K\"{o}nigl}(1979)}]{BK79}
{Blandford}, R.~D., \& {K\"{o}nigl}, A. 1979, \apj, 232, 34,
  \dodoi{10.1086/157262}

\bibitem[{{Blandford} \& {Znajek}(1977)}]{BZ77}
{Blandford}, R.~D., \& {Znajek}, R.~L. 1977, \mnras, 179, 433,
  \dodoi{10.1093/mnras/179.3.433}

\bibitem[{{Comisso} \& {Sironi}(2018)}]{Comisso18}
{Comisso}, L., \& {Sironi}, L. 2018, \prl, 121, 255101,
  \dodoi{10.1103/PhysRevLett.121.255101}

\bibitem[{{Comisso} \& {Sironi}(2019)}]{Comisso19}
---. 2019, \apj, 886, 122, \dodoi{10.3847/1538-4357/ab4c33}

\bibitem[{{Crumley} {et~al.}(2019){Crumley}, {Caprioli}, {Markoff}, \&
  {Spitkovsky}}]{Crumley19}
{Crumley}, P., {Caprioli}, D., {Markoff}, S., \& {Spitkovsky}, A. 2019, \mnras,
  485, 5105, \dodoi{10.1093/mnras/stz232}

\bibitem[{{Davis} \& {Tchekhovskoy}(2020)}]{Davis20}
{Davis}, S.~W., \& {Tchekhovskoy}, A. 2020, \araa, 58, 407,
  \dodoi{10.1146/annurev-astro-081817-051905}

\bibitem[{{Ghisellini}(2012)}]{Ghisellini12}
{Ghisellini}, G. 2012, \mnras, 424, L26,
  \dodoi{10.1111/j.1745-3933.2012.01280.x}

\bibitem[{{Giommi} {et~al.}(2012){Giommi}, {Polenta}, {L{\"a}hteenm{\"a}ki},
  {Thompson}, {Capalbi}, {Cutini}, {Gasparrini}, {Gonz{\'a}lez-Nuevo},
  {Le{\'o}n-Tavares}, {L{\'o}pez-Caniego}, {Mazziotta}, {Monte}, {Perri},
  {Rain{\`o}}, {Tosti}, {Tramacere}, {Verrecchia}, {Aller}, {Aller},
  {Angelakis}, {Bastieri}, {Berdyugin}, {Bonaldi}, {Bonavera}, {Burigana},
  {Burrows}, {Buson}, {Cavazzuti}, {Chincarini}, {Colafrancesco}, {Costamante},
  {Cuttaia}, {D'Ammando}, {de Zotti}, {Frailis}, {Fuhrmann}, {Galeotta},
  {Gargano}, {Gehrels}, {Giglietto}, {Giordano}, {Giroletti}, {Keih{\"a}nen},
  {King}, {Krichbaum}, {Lasenby}, {Lavonen}, {Lawrence}, {Leto}, {Lindfors},
  {Mandolesi}, {Massardi}, {Max-Moerbeck}, {Michelson}, {Mingaliev}, {Natoli},
  {Nestoras}, {Nieppola}, {Nilsson}, {Partridge}, {Pavlidou}, {Pearson},
  {Procopio}, {Rachen}, {Readhead}, {Reeves}, {Reimer}, {Reinthal},
  {Ricciardi}, {Richards}, {Riquelme}, {Saarinen}, {Sajina}, {Sandri},
  {Savolainen}, {Sievers}, {Sillanp{\"a}{\"a}}, {Sotnikova}, {Stevenson},
  {Tagliaferri}, {Takalo}, {Tammi}, {Tavagnacco}, {Terenzi}, {Toffolatti},
  {Tornikoski}, {Trigilio}, {Turunen}, {Umana}, {Ungerechts}, {Villa}, {Wu},
  {Zacchei}, {Zensus}, \& {Zhou}}]{Giommi12}
{Giommi}, P., {Polenta}, G., {L{\"a}hteenm{\"a}ki}, A., {et~al.} 2012, \aap,
  541, A160, \dodoi{10.1051/0004-6361/201117825}

\bibitem[{{Goldreich} \& {Julian}(1969)}]{GJ69}
{Goldreich}, P., \& {Julian}, W.~H. 1969, \apj, 157, 869,
  \dodoi{10.1086/150119}

\bibitem[{{Grier} {et~al.}(2017){Grier}, {Pancoast}, {Barth}, {Fausnaugh},
  {Brewer}, {Treu}, \& {Peterson}}]{Grier17}
{Grier}, C.~J., {Pancoast}, A., {Barth}, A.~J., {et~al.} 2017, \apj, 849, 146,
  \dodoi{10.3847/1538-4357/aa901b}

\bibitem[{{Henri} \& {Pelletier}(1991)}]{Henri91}
{Henri}, G., \& {Pelletier}, G. 1991, \apjl, 383, L7, \dodoi{10.1086/186228}

\bibitem[{{Janiak} {et~al.}(2016){Janiak}, {Sikora}, \& {Moderski}}]{Janiak16}
{Janiak}, M., {Sikora}, M., \& {Moderski}, R. 2016, \mnras, 458, 2360,
  \dodoi{10.1093/mnras/stw465}

\bibitem[{{Johnson} {et~al.}(1993){Johnson}, {Kinzer}, {Kurfess}, {Strickman},
  {Purcell}, {Grabelsky}, {Ulmer}, {Hillis}, {Jung}, \& {Cameron}}]{Johnson93}
{Johnson}, W.~N., {Kinzer}, R.~L., {Kurfess}, J.~D., {et~al.} 1993, \apjs, 86,
  693, \dodoi{10.1086/191795}

\bibitem[{{Jorstad} {et~al.}(2005){Jorstad}, {Marscher}, {Lister}, {Stirling},
  {Cawthorne}, {Gear}, {G{\'o}mez}, {Stevens}, {Smith}, {Forster}, \&
  {Robson}}]{Jorstad05}
{Jorstad}, S.~G., {Marscher}, A.~P., {Lister}, M.~L., {et~al.} 2005, \aj, 130,
  1418, \dodoi{10.1086/444593}

\bibitem[{{Jorstad} {et~al.}(2017){Jorstad}, {Marscher}, {Morozova},
  {Troitsky}, {Agudo}, {Casadio}, {Foord}, {G{\'o}mez}, {MacDonald}, {Molina},
  {L{\"a}hteenm{\"a}ki}, {Tammi}, \& {Tornikoski}}]{Jorstad17}
{Jorstad}, S.~G., {Marscher}, A.~P., {Morozova}, D.~A., {et~al.} 2017, \apj,
  846, 98, \dodoi{10.3847/1538-4357/aa8407}

\bibitem[{{Kang} {et~al.}(2014){Kang}, {Chen}, \& {Wu}}]{Kang14}
{Kang}, S.-J., {Chen}, L., \& {Wu}, Q. 2014, \apjs, 215, 5,
  \dodoi{10.1088/0067-0049/215/1/5}

\bibitem[{{Kataoka} {et~al.}(2011){Kataoka}, {Stawarz}, {Takahashi}, {Cheung},
  {Hayashida}, {Grandi}, {Burnett}, {Celotti}, {Fegan}, {Fortin}, {Maeda},
  {Nakamori}, {Taylor}, {Tosti}, {Digel}, {McConville}, {Finke}, \&
  {D'Ammando}}]{Kataoka11}
{Kataoka}, J., {Stawarz}, {\L}., {Takahashi}, Y., {et~al.} 2011, \apj, 740, 29,
  \dodoi{10.1088/0004-637X/740/1/29}

\bibitem[{{Komissarov} {et~al.}(2009){Komissarov}, {Vlahakis}, {K{\"o}nigl}, \&
  {Barkov}}]{Komissarov09}
{Komissarov}, S.~S., {Vlahakis}, N., {K{\"o}nigl}, A., \& {Barkov}, M.~V. 2009,
  \mnras, 394, 1182, \dodoi{10.1111/j.1365-2966.2009.14410.x}

\bibitem[{{K\"{o}nigl}(1981)}]{Konigl81}
{K\"{o}nigl}, A. 1981, \apj, 243, 700, \dodoi{10.1086/158638}

\bibitem[{{Kovalev} {et~al.}(2008){Kovalev}, {Lobanov}, {Pushkarev}, \&
  {Zensus}}]{Kovalev08}
{Kovalev}, Y.~Y., {Lobanov}, A.~P., {Pushkarev}, A.~B., \& {Zensus}, J.~A.
  2008, \aap, 483, 759, \dodoi{10.1051/0004-6361:20078679}

\bibitem[{{Kovalev} {et~al.}(2020){Kovalev}, {Pushkarev}, {Nokhrina}, {Plavin},
  {Beskin}, {Chernoglazov}, {Lister}, \& {Savolainen}}]{Kovalev20}
{Kovalev}, Y.~Y., {Pushkarev}, A.~B., {Nokhrina}, E.~E., {et~al.} 2020, \mnras,
  495, 3576, \dodoi{10.1093/mnras/staa1121}

\bibitem[{{Levinson} \& {Rieger}(2011)}]{Levinson11}
{Levinson}, A., \& {Rieger}, F. 2011, \apj, 730, 123,
  \dodoi{10.1088/0004-637X/730/2/123}

\bibitem[{{Liodakis} {et~al.}(2022){Liodakis}, {Blinov}, {Potter}, \&
  {Rieger}}]{Liodakis22}
{Liodakis}, I., {Blinov}, D., {Potter}, S.~B., \& {Rieger}, F.~M. 2022, \mnras,
  509, L21, \dodoi{10.1093/mnrasl/slab118}

\bibitem[{{Lobanov}(1998)}]{Lobanov98}
{Lobanov}, A.~P. 1998, \aap, 330, 79.
\newblock \doarXiv{astro-ph/9712132}

\bibitem[{{Lohfink} {et~al.}(2013){Lohfink}, {Reynolds}, {Jorstad}, {Marscher},
  {Miller}, {Aller}, {Aller}, {Brenneman}, {Fabian}, {Miller}, {Mushotzky},
  {Nowak}, \& {Tombesi}}]{Lohfink13}
{Lohfink}, A.~M., {Reynolds}, C.~S., {Jorstad}, S.~G., {et~al.} 2013, \apj,
  772, 83, \dodoi{10.1088/0004-637X/772/2/83}

\bibitem[{{Madejski} \& {Sikora}(2016)}]{MS16}
{Madejski}, G.~G., \& {Sikora}, M. 2016, \araa, 54, 725,
  \dodoi{10.1146/annurev-astro-081913-040044}

\bibitem[{{Marscher} {et~al.}(2002){Marscher}, {Jorstad}, {G{\'o}mez}, {Aller},
  {Ter{\"a}sranta}, {Lister}, \& {Stirling}}]{Marscher02_3C120}
{Marscher}, A.~P., {Jorstad}, S.~G., {G{\'o}mez}, J.-L., {et~al.} 2002, \nat,
  417, 625, \dodoi{10.1038/nature00772}

\bibitem[{{McKinney} {et~al.}(2012){McKinney}, {Tchekhovskoy}, \&
  {Blandford}}]{McKinney12}
{McKinney}, J.~C., {Tchekhovskoy}, A., \& {Blandford}, R.~D. 2012, \mnras, 423,
  3083, \dodoi{10.1111/j.1365-2966.2012.21074.x}

\bibitem[{{Mo{\'s}cibrodzka} {et~al.}(2011){Mo{\'s}cibrodzka}, {Gammie},
  {Dolence}, \& {Shiokawa}}]{Moscibrodzka11}
{Mo{\'s}cibrodzka}, M., {Gammie}, C.~F., {Dolence}, J.~C., \& {Shiokawa}, H.
  2011, \apj, 735, 9, \dodoi{10.1088/0004-637X/735/1/9}

\bibitem[{{Narayan} {et~al.}(2003){Narayan}, {Igumenshchev}, \&
  {Abramowicz}}]{Narayan03}
{Narayan}, R., {Igumenshchev}, I.~V., \& {Abramowicz}, M.~A. 2003, \pasj, 55,
  L69, \dodoi{10.1093/pasj/55.6.L69}

\bibitem[{{Nokhrina} {et~al.}(2015){Nokhrina}, {Beskin}, {Kovalev}, \&
  {Zheltoukhov}}]{Nokhrina15}
{Nokhrina}, E.~E., {Beskin}, V.~S., {Kovalev}, Y.~Y., \& {Zheltoukhov}, A.~A.
  2015, \mnras, 447, 2726, \dodoi{10.1093/mnras/stu2587}

\bibitem[{{Ogle} {et~al.}(2005){Ogle}, {Davis}, {Antonucci}, {Colbert},
  {Malkan}, {Page}, {Sasseen}, \& {Tornikoski}}]{Ogle05}
{Ogle}, P.~M., {Davis}, S.~W., {Antonucci}, R.~R.~J., {et~al.} 2005, \apj, 618,
  139, \dodoi{10.1086/425894}

\bibitem[{{Oh} {et~al.}(2018){Oh}, {Koss}, {Markwardt}, {Schawinski},
  {Baumgartner}, {Barthelmy}, {Cenko}, {Gehrels}, {Mushotzky}, {Petulante},
  {Ricci}, {Lien}, \& {Trakhtenbrot}}]{Oh18}
{Oh}, K., {Koss}, M., {Markwardt}, C.~B., {et~al.} 2018, \apjs, 235, 4,
  \dodoi{10.3847/1538-4365/aaa7fd}

\bibitem[{{Petropoulou} {et~al.}(2019){Petropoulou}, {Sironi}, {Spitkovsky}, \&
  {Giannios}}]{Petropoulou19}
{Petropoulou}, M., {Sironi}, L., {Spitkovsky}, A., \& {Giannios}, D. 2019,
  \apj, 880, 37, \dodoi{10.3847/1538-4357/ab287a}

\bibitem[{{Pjanka} {et~al.}(2017){Pjanka}, {Zdziarski}, \& {Sikora}}]{Pjanka17}
{Pjanka}, P., {Zdziarski}, A.~A., \& {Sikora}, M. 2017, \mnras, 465, 3506,
  \dodoi{10.1093/mnras/stw2960}

\bibitem[{{Punsly} {et~al.}(2009){Punsly}, {Igumenshchev}, \&
  {Hirose}}]{Punsly09}
{Punsly}, B., {Igumenshchev}, I.~V., \& {Hirose}, S. 2009, \apj, 704, 1065,
  \dodoi{10.1088/0004-637X/704/2/1065}

\bibitem[{{Pushkarev} {et~al.}(2017){Pushkarev}, {Kovalev}, {Lister}, \&
  {Savolainen}}]{Pushkarev17}
{Pushkarev}, A.~B., {Kovalev}, Y.~Y., {Lister}, M.~L., \& {Savolainen}, T.
  2017, \mnras, 468, 4992, \dodoi{10.1093/mnras/stx854}

\bibitem[{{Richards} {et~al.}(2006){Richards}, {Lacy}, {Storrie-Lombardi},
  {Hall}, {Gallagher}, {Hines}, {Fan}, {Papovich}, {Vanden Berk}, {Trammell},
  {Schneider}, {Vestergaard}, {York}, {Jester}, {Anderson}, {Budav{\'a}ri}, \&
  {Szalay}}]{Richards06}
{Richards}, G.~T., {Lacy}, M., {Storrie-Lombardi}, L.~J., {et~al.} 2006, \apjs,
  166, 470, \dodoi{10.1086/506525}

\bibitem[{{Sikora}(2016)}]{Sikora16}
{Sikora}, M. 2016, Galaxies, 4, 12, \dodoi{10.3390/galaxies4030012}

\bibitem[{{Sikora} {et~al.}(1997){Sikora}, {Madejski}, {Moderski}, \&
  {Poutanen}}]{Sikora97}
{Sikora}, M., {Madejski}, G., {Moderski}, R., \& {Poutanen}, J. 1997, \apj,
  484, 108, \dodoi{10.1086/304305}

\bibitem[{{Sikora} {et~al.}(2020){Sikora}, {Nalewajko}, \&
  {Madejski}}]{Sikora20}
{Sikora}, M., {Nalewajko}, K., \& {Madejski}, G.~M. 2020, \mnras, 499, 3749,
  \dodoi{10.1093/mnras/staa3128}

\bibitem[{{Sironi} {et~al.}(2015){Sironi}, {Petropoulou}, \&
  {Giannios}}]{Sironi15b}
{Sironi}, L., {Petropoulou}, M., \& {Giannios}, D. 2015, \mnras, 450, 183,
  \dodoi{10.1093/mnras/stv641}

\bibitem[{{Sironi} \& {Spitkovsky}(2009)}]{Sironi09}
{Sironi}, L., \& {Spitkovsky}, A. 2009, \apj, 698, 1523,
  \dodoi{10.1088/0004-637X/698/2/1523}

\bibitem[{{Snios} {et~al.}(2018){Snios}, {Nulsen}, {Wise}, {de Vries},
  {Birkinshaw}, {Worrall}, {Duffy}, {Kraft}, {McNamara}, {Carilli}, {Croston},
  {Edge}, {Godfrey}, {Hardcastle}, {Harris}, {Laing}, {Mathews}, {McKean},
  {Perley}, {Rafferty}, \& {Young}}]{Snios18}
{Snios}, B., {Nulsen}, P. E.~J., {Wise}, M.~W., {et~al.} 2018, \apj, 855, 71,
  \dodoi{10.3847/1538-4357/aaaf1a}

\bibitem[{{Svensson}(1987)}]{Svensson87}
{Svensson}, R. 1987, \mnras, 227, 403

\bibitem[{{Tchekhovskoy} {et~al.}(2009){Tchekhovskoy}, {McKinney}, \&
  {Narayan}}]{Tchekhovskoy09}
{Tchekhovskoy}, A., {McKinney}, J.~C., \& {Narayan}, R. 2009, \apj, 699, 1789,
  \dodoi{10.1088/0004-637X/699/2/1789}

\bibitem[{{Tchekhovskoy} {et~al.}(2011){Tchekhovskoy}, {Narayan}, \&
  {McKinney}}]{Tchekhovskoy11}
{Tchekhovskoy}, A., {Narayan}, R., \& {McKinney}, J.~C. 2011, \mnras, 418, L79,
  \dodoi{10.1111/j.1745-3933.2011.01147.x}

\bibitem[{{Wo{\'z}niak} {et~al.}(1998){Wo{\'z}niak}, {Zdziarski}, {Smith},
  {Madejski}, \& {Johnson}}]{Wozniak98}
{Wo{\'z}niak}, P.~R., {Zdziarski}, A.~A., {Smith}, D., {Madejski}, G.~M., \&
  {Johnson}, W.~N. 1998, \mnras, 299, 449,
  \dodoi{10.1046/j.1365-8711.1998.01831.x}

\bibitem[{{Zamaninasab} {et~al.}(2014){Zamaninasab}, {Clausen-Brown},
  {Savolainen}, \& {Tchekhovskoy}}]{Zamaninasab14}
{Zamaninasab}, M., {Clausen-Brown}, E., {Savolainen}, T., \& {Tchekhovskoy}, A.
  2014, \nat, 510, 126, \dodoi{10.1038/nature13399}

\bibitem[{{Zdziarski} \& {Grandi}(2001)}]{ZG01}
{Zdziarski}, A.~A., \& {Grandi}, P. 2001, \apj, 551, 186,
  \dodoi{10.1086/320064}

\bibitem[{{Zdziarski} {et~al.}(2015){Zdziarski}, {Sikora}, {Pjanka}, \&
  {Tchekhovskoy}}]{ZSPT15}
{Zdziarski}, A.~A., {Sikora}, M., {Pjanka}, P., \& {Tchekhovskoy}, A. 2015,
  \mnras, 451, 927, \dodoi{10.1093/mnras/stv986}

\bibitem[{{Zdziarski} {et~al.}(2019){Zdziarski}, {Stawarz}, \&
  {Sikora}}]{Zdziarski19b}
{Zdziarski}, A.~A., {Stawarz}, {\L}., \& {Sikora}, M. 2019, \mnras, 485, 1210,
  \dodoi{10.1093/mnras/stz475}

\bibitem[{{Zdziarski} {et~al.}(2022){Zdziarski}, {Tetarenko}, \&
  {Sikora}}]{Zdziarski22}
{Zdziarski}, A.~A., {Tetarenko}, A.~J., \& {Sikora}, M. 2022, \apj, 925, 189,
  \dodoi{10.3847/1538-4357/ac38a9}

\bibitem[{{Zdziarski} {et~al.}(2021){Zdziarski}, {Jourdain}, {Lubi{\'n}ski},
  {Szanecki}, {Nied{\'z}wiecki}, {Veledina}, {Poutanen}, {Dzie{\l}ak}, \&
  {Roques}}]{Zdziarski21c}
{Zdziarski}, A.~A., {Jourdain}, E., {Lubi{\'n}ski}, P., {et~al.} 2021, \apjl,
  914, L5, \dodoi{10.3847/2041-8213/ac0147}

\end{thebibliography}
\bibliographystyle{aasjournal}

\end{document}